\documentclass{tdp-clean}
\usepackage[utf8]{inputenc}
\usepackage[inline]{enumitem}
\usepackage{amsmath}
\usepackage{url}

\begin{document}

\title{How to Get Actual Privacy and Utility from Privacy Models: the $k$-Anonymity and Differential Privacy Families}
\author{Josep Domingo-Ferrer, David S\'anchez}
\address{Department of Computer Science and Mathematics, CYBERCAT-Center for Cybersecurity Research of Catalonia, Universitat Rovira i Virgili, Av. Pa\"{\i}sos Catalans, 26, Tarragona, 43007, Catalonia. \\
  E-mail: {\small \tt{\{josep.domingo,david.sanchez\}@urv.cat}}
}


\maketitle

\begin{abstract}
Privacy models were introduced in the
privacy-preserving data publishing and statistical disclosure control literature with the promise to end the need for costly empirical assessment of disclosure risk.  
We examine how well this promise is kept by the main privacy models. 
We find they may fail to provide adequate protection guarantees because of problems in their definition
or incur unacceptable trade-offs between privacy protection and utility preservation. Specifically, $k$-anonymity may not entirely exclude disclosure 
if enforced with deterministic mechanisms or without constraints on the confidential values.
On the other hand, differential privacy (DP) incurs unacceptable utility loss for small budgets and its privacy guarantee becomes meaningless for large budgets. In the latter case, an \emph{ex post} empirical assessment of disclosure risk becomes necessary, undermining the main appeal of privacy models. 
Whereas the utility preservation of DP can only be improved by relaxing its privacy guarantees, 
we argue that a semantic reformulation of $k$-anonymity can offer more robust privacy  
without losing utility with respect to traditional syntactic $k$-anonymity. 
\end{abstract}

\begin{keywords}
Data privacy, privacy models, $k$-anonymity, differential privacy, probabilistic $k$-anonymity
\end{keywords}

\section{Introduction}
Since the 1960s, research has been conducted with the goal of collecting and publishing analytically useful data that cannot be linked to specific individuals or
reveal confidential information about them. Over the years, a substantial body of privacy-enhancing methods has been developed in the literature on statistical disclosure control (SDC) and privacy-preserving data publishing (PPDP).

The traditional approach to privacy protection in official statistics is heuristic: an SDC method, such as data suppression, sampling, or noise addition, is iteratively tried with varying parameter values until a parameter choice is found that preserves sufficient analytical utility while bringing the disclosure risk below a certain threshold. In this approach, both utility and disclosure risk are evaluated \emph{ex post} by empirically measuring information loss and the probability of disclosure in actual protected outputs. 

In the late 1990s, due to the surge of the web and the consequent expansion of data collection and exchange beyond official statistics, privacy protection became a mainstream topic in the computer science community, which approached it under a different angle. 
Specifically, an \emph{ex ante} privacy condition is 
stated, which is then enforced on the actual data to be protected by using one or several SDC methods.
Since privacy is modeled as the satisfaction of a certain condition, this approach is commonly referred to as \emph{privacy model}. A significant advantage of privacy models with respect to heuristic SDC methods is that, since the former enforce an \emph{ex ante} privacy condition, the \emph{ex post} empirical evaluation of
disclosure risk may no longer be necessary. In other words, privacy models are supposed to \emph{guarantee} a certain level of privacy \emph{by design}. 

The first privacy model was $k$-anonymity \cite{kanonymity,k-anonymity}, which was followed by extensions such as $l$-diversity \cite{diversity} and $t$-closeness \cite{tcloseness}. A substantially different family of privacy models was inaugurated by $\epsilon$-differential privacy (DP) \cite{dwork2006} and subsequently expanded by a variety of relaxations \cite{pejo2020sok}. This second family is by far the most widespread these days.

In this paper, we critically review these two families of privacy models and discuss the privacy guarantees they actually provide. 
In its traditional syntactic formulation, the $k$-anonymity family may fail to guarantee privacy in very specific cases. 
On the other hand, the privacy guarantee of the DP family might be too broad for most applications and become meaningless unless the budget $\epsilon$ is small, which comes at the cost of unacceptable utility loss. 
Whereas the preservation of utility by DP can only be improved by relaxing its privacy guarantee, we argue that a semantic reformulation of $k$-anonymity can 
offer more robust guarantees while retaining its favorable and
selectable privacy-utility trade-off.

The remainder of this paper is structured as follows. Privacy notions, including the types of disclosure risks and the types of data attributes, are reviewed in Section~\ref{sec:notions}. 
Section~\ref{sec:models} gives an overview of privacy models and their privacy promises. Objections and mitigations to $k$-anonymity and DP are discussed in Sections~\ref{sec:kanon} and~\ref{sec:dp}, respectively. Conclusions are
drawn in Section~\ref{sec:conclusion}.

\section{Privacy notions}
\label{sec:notions}

To understand the guarantees offered by privacy models, we first need to consider the types of information disclosure:
\begin{itemize}
\item \emph{Identity disclosure} occurs when an adversary can identify the subject to whom a released output corresponds. This is known as \emph{reidentification}.
\item \emph{Attribute disclosure} occurs when seeing the released outputs allows the adversary to determine or
closely approximate the value of a subject’s confidential attribute.
\item \emph{Membership disclosure} occurs when the adversary can determine whether a certain subject’s data 
were part of the data used to compute the released outputs.
\end{itemize}

Since revealing confidential attributes ({\em e.g.}, income, health condition, etc.) causes a serious privacy threat, \emph{attribute disclosure is arguably the most critical type of disclosure}. 
From that point of view, reidentification and membership disclosure only result in disclosure of privacy if \emph{they enable attribute disclosure}. This will happen when reidentified records (\emph{e.g.}, patient records) can be unequivocally associated with confidential attributes (\emph{e.g.}, health conditions), or when inferring membership (\emph{e.g.}, a patient being a member of a released data set) implicitly reveals confidential values (\emph{e.g.}, if it is known that all patients in the released data set suffer from the same condition). Moreover, whereas reidentification allows singling out an individual's record in a released data set, membership disclosure becomes irrelevant when the published data are an exhaustive representation of a population ({\em e.g.}, a national census) because there is no uncertainty about the membership of any subject. 

Identity and attribute disclosures have been considered in the privacy literature for decades~\cite{sdc-book}, whereas membership disclosure has only gained relevance in the last few years within the
machine learning field~\cite{shokri2017membership}.
The reason is that, quite often, the only plausible attack against black-box machine learning models is membership inference.

In line with the above three types of disclosure, attributes in a data release can be classified into different types.
\emph{Identifiers} are attributes that allow unambiguous reidentification of the individual to whom they refer. Examples include social security numbers or passport numbers. 
Identifiers should always be suppressed in anonymized data releases.
{\em Quasi-identifiers} (QI) are attributes that do not lead
to re-reidentification when taken one by one, but a combination of them may if i) it is known by the intruder on a target individual and ii) is unique in the population.
For example, \cite{Sweeney2000} shows that 87\% of the population in the United States can be unambiguously identified once their 5-digit ZIP code, birth date, and sex are known.
{\em Confidential attributes} hold confidential
information on the individual (\emph{e.g.}, salary, health condition, sexual orientation, etc.). As stated above, the primary goal of privacy protection is to prevent
intruders from determining or closely approximating confidential information about a specific
individual. \emph{Non-confidential attributes} are those that do not belong to any of the previous categories. As they do not contain confidential information about individuals and cannot be used for record reidentification, they do not need to be taken into account when protecting privacy. 

\section{Overview of privacy models}
\label{sec:models}
Given an original sensitive data set $D$ and an anonymized output $Y$ computed on $D$ in a prescribed manner, a privacy model $M$ with parameters $\lambda$, that is, $M(D,Y,\lambda)$ is a mathematical guarantee in terms of $\lambda$ that limits at least one of the three disclosure risks (identity, attribute, or membership) associated with the release of $Y$.

Beyond offering \emph{ex ante} guarantees, a model $M$ may exhibit additional desirable properties:
\begin{enumerate}
    \item \label{immunity} {\em Immunity to post-processing:}  
    If a mechanism \(\mathcal{A}\) satisfies $M$ with parameters $\lambda$, any post-processing function \(g(\cdot)\) applied to its output also satisfies $M$ with parameters $\lambda$.
    \item {\em Sequential composition:} If a mechanism \(\mathcal{A}_1\) satisfies $M$ with parameters $\lambda_1$, and the mechanism \(\mathcal{A}_2\) satisfies $M$ with parameters $\lambda_2$, their combined application $\mathcal{A}_{\text{sequential}}$ on the same data set or non-disjoint data sets satisfies $M$ with weaker parameters $lambda_3=f(lambda_1,lamba_2)$.
    \item {\em Parallel composition:} If the mechanisms \(\mathcal{A}_1\) and \(\mathcal{A}_2\) both satisfy $M$ with parameter $\lambda$ and operate on disjoint data sets, their combined application $\mathcal{A}_{\text{parallel}}$ satisfies $M$ with the same parameters $\lambda$.
\end{enumerate}

\subsection{$k$-Anonymity and extensions}
\label{kanonymity}

$k$-Anonymity \cite{kanonymity} is a privacy model specifically designed to anonymize outputs consisting of microdata releases, that is, data sets of records with detailed attribute values related to specific individuals.

$k$-Anonymity assumes that the values of the quasi-identifier attributes are known to the intruder,
who can use them for reidentification. Anonymized microdata must be computed so that each combination of values of the quasi-identifier
attributes is shared by $k$ or more records, which form a so-called $k$-anonymous class. This makes records indistinguishable (and in principle not reidentifiable without ambiguity) within each class. Therefore, the mathematical guarantee that can be given is an upper bound $1/k$ on the risk of reidentification of each individual record. 

$k$-Anonymity is not tied to a specific masking mechanism, and can be achieved by masking quasi-identifiers through a variety of SDC methods, including generalization/suppression (the original approach in \cite{kanonymity}) or aggregation (the microaggregation approach in \cite{micro}).

Although plain $k$-anonymity can protect against reidentification, attribute disclosure may still occur if the values of a confidential attribute within a $k$-anonymous class are the same or very similar.
In this case, the intruder just needs to find out that their target record belongs to that $k$-anonymous class to infer the target's confidential attribute value; that is, attribute disclosure occurs without prior reidentification. To address this issue, $k$-anonymity extensions were proposed. 

The first extension was $l$-diversity \cite{diversity}, in which anonymized microdata must be calculated so that there are 
$l$ ``well-represented'' values for each confidential attribute within each $k$-anonymous class. The meaning of ``well-represented'' is open to interpretation and can be understood as \emph{different} values or as values sampled from a distribution whose Shannon entropy is above a certain threshold. 
Given a criterion of well-representedness, the higher $l$, the higher the intruder's uncertainty about the value of any confidential attribute for his/her target record.   

While $l$-diversity defines an absolute measure of variability for confidential values, $t$-closeness~\cite{tcloseness} is another extension of $k$-anonymity whose aim is to limit the gain of information on the confidential attribute in case the intruder manages to locate the $k$-anonymous class of their target record. The gain is measured w.r.t. the marginal distribution of the confidential attribute in the entire data set, which is assumed known to the intruder. $t$-Closeness requires the anonymized microdata set to be calculated so that the distance between the distribution of each confidential attribute in any $k$-anonymous class and the distribution of the attribute in the entire data set is not more than a threshold $t$. The smaller $t$, the more limited the intruder's gain. A common distance measure to enforce $t$-closeness is the earth mover's distance.  

\subsection{$\epsilon$-Differential privacy and relaxations}
\label{dp}

$\epsilon$-Differential privacy (DP) was originally proposed in~\cite{dwork2006} as
a privacy model designed to protect the outputs of interactive queries on an unreleased sensitive data set. The principle underlying DP is that
the presence or absence of any single individual's record in the data set should be indistinguishable
on the basis of the released query answer. 

More formally, a randomized function $\kappa$ gives $\epsilon$-differential privacy if, for all neighbor data sets $X_{1}$ and $X_{2}$ that differ in one record and all $S\subset Range(\kappa)$, we have
\[
\label{eq:dp}
\Pr(\kappa(X_{1})\in S)\le\exp(\epsilon)\times\Pr(\kappa(X_{2})\in S).
\]

The exponential parameter $\epsilon$ (also called privacy budget) defines how indistinguishable the presence or absence of any single record is based on the query answers: the smaller $\epsilon$, the higher the protection. 
Hence, the anonymized query output must be computed using \emph{randomization}, and the mathematical guarantee offered by DP is an upper bound in terms of the parameter $\epsilon$ on the risk of {\em membership disclosure} for any individual who contributed to the unreleased original data set. 

DP is commonly enforced by adding Laplacian noise that is inversely proportional to $\epsilon$ and directly proportional to the global sensitivity of the query, that is, how much the output of the (unrandomized) query can change due to the addition, deletion, or change of a single record in the data set. Mild noise may suffice for aggregated statistical queries, such as the mean, but much larger noise is needed for more sensitive queries, like 
maximum, minimum, or identity queries ({\em i.e.}, querying the value of the $i$-th record).

Even if DP was designed to protect interactive queries, it has been extensively applied to non-interactive scenarios such as continuous data collection and (micro)data releases \cite{Blanco2023,practice}. However, applying DP to record-level data (as is the case for data collection or microdata release) is equivalent to
answering identity queries that, because of the large sensitivity of any individual's data, require adding a large amount of noise to attain a ``safe'' $\epsilon$. This severely hampers the utility of the DP-protected outputs \cite{VLDB}.

Due to the impossibility of reconciling data utility with small enough $\epsilon$ in non-interactive settings, a variety of relaxations of the DP definition have been proposed. The most widely used are:

\begin{itemize}
    \item $(\epsilon, \delta)$-DP ~\cite{dwork2006}, where parameter $\delta$ is the probability that pure DP is not satisfied.     
    \item \emph{R\'enyi DP} (RDP) \cite{mironovRenyi2017} is a relaxation of DP based on the R\'enyi divergence through parameter $\alpha \in (1, \infty)$, providing a quantifiable measure of privacy leakage. When $\alpha=1$, RDP converges to the Kullback-Leibler (KL) divergence and, for $\alpha=\infty$, it converges to pure DP.
    \item \emph{Zero-Concentrated DP} (zCDP) \cite{cdp216} is also defined in terms of the R\'enyi divergence. However, a single parameter $\rho \in (1, \infty)$ restricts parameter $\alpha$.  For $\rho=1$, zCDP is equivalent to the KL divergence and, for $\rho=\infty$, it approaches the max-divergence and pure DP.    
\end{itemize}

\section{Objections and mitigations to $k$-anonymity}
\label{sec:kanon}

$k$-Anonymity is designed to protect against \emph{identity disclosure} and,
unlike DP, offers a
\emph{clear and meaningful guarantee for any value of its parameter}. Specifically, for any $k>1$, the probability of successful reidentification should be at most $1/k$,
which should eliminate the need to empirically evaluate
the risk of reidentification \emph{ex post} for any value of $k$.

However, $k$-anonymity is often dismissed and considered to be more obsolete than DP due to alleged flaws, which we will discuss next. For each of them, we highlight solutions or mitigations.

\subsection{Formal problems of the $k$-anonymity definition}
\label{formal}

The need to make assumptions on the list of attributes that can be used as quasi-identifiers by an intruder has often been pointed out as a weakness of $k$-anonymity. A way to circumvent the need for such assumptions is to consider \emph{all} attributes as potential quasi-identifiers. Obviously, the more attributes are masked to enforce $k$-anonymity, the less utility is preserved by the anonymized microdata set. However, masking all attributes is exactly what DP also does when used to generate anonymized microdata. 
In this context, $k$-anonymity is more flexible and consistent than DP, for several reasons:
\begin{enumerate}
    \item Better utility can be attained with $k$-anonymity if one can make realistic assumptions on the attributes that can or cannot be plausibly used by an intruder as quasi-identifiers. 
    \item For $k$-anonymity to be fulfilled, it is enough to mask quasi-identifier values only in those original records whose quasi-identifier combination has frequency less than $k$, rather than masking all attribute values in all records as DP does. 
    \item $k$-Anonymity ensures that \emph{all original records at risk} ({\em i.e.}, with reidentification risk above $1/k$) are modified for any value of $k$; 
    in contrast, DP with large $\epsilon$ may leave many records at risk virtually untouched or with extremely small noise.
    \end{enumerate}

A more recent criticism stems from the \emph{syntactic} formulation of $k$-anonymity, which focuses on the structure of the data, rather than on their meaning. Indeed, the requirement of having at least $k$ records sharing the same combination of quasi-identifiers can be achieved with \emph{deterministic} mechanisms that are vulnerable to attack. Specifically, a data set can be made $k$-anonymous by replacing attribute values of non-$k$-anonymous records by common generalizations (\emph{e.g.}, $\{USA, France\} \rightarrow \{Country\}$). However, when generalizations fulfill \emph{minimality}, which means that no record is generalized more than necessary to achieve the $k$-anonymity property, and the hierarchy of generalizations is known, a recent study \cite{Cohen} shows that it might be possible to \emph{downcode} (\emph{i.e.}, undo) some generalizations because the minimal generalizations of specific data distributions are deterministic and (partially) reversible. 

Although \cite{Cohen} hints that this vulnerability affects all approaches to $k$-anonymity, in reality it only applies to $k$-anonymity via minimum generalization, which is very rarely used. Actually, any mechanism
to achieve $k$-anonymity that is not reversible can withstand downcoding:
\begin{itemize}
    \item If generalization is used, although minimum generalization is the least utility-damaging, it is computationally hard to find it. In practice, $k$-anonymity via generalization is achieved via suboptimal heuristics that yield non-minimum generalizations --that are not reversible--. Additionally, non-minimum generalization is often combined with suppression to limit information loss, and suppression is not reversible~\cite{k-anonymity,sweeney2002achieving}.
    \item Averaging the quasi-identifier values within each $k$-anonymous class is another usual mechanism for $k$-anonymity that is not reversible~\cite{micro}. 
\end{itemize}

In any case, it is true that the syntactic formulation of $k$-anonymity does not exclude the use of a mechanism such as minimum generalization that is vulnerable to \emph{downcoding} attacks. To address this issue, we suggest employing a {\em semantic} formulation that focuses on the meaning of the data, rather than on their structure. Specifically, probabilistic $k$-anonymity
eliminates the standard $k$-anonymity requirement that any combination of quasi-identifiers be replicated $k$ or more times and makes the privacy guarantee explicit, as follows \cite{probabilistic}.

A data set $\mathcal{D}'$ generated from an original data set $\mathcal{D}$ via mechanism $\mathcal{A}$ is said to satisfy {\em probabilistic $k$-anonymity} if, for any non-anonymous external data set $\mathcal{E}$, the probability that an intruder knowing $\mathcal{D}'$, $\mathcal{A}$ and $\mathcal{E}$ correctly links any record in $\mathcal{E}$ with its corresponding record (if any) in $\mathcal{D}'$ is at most $1/k$.

Since probabilistic $k$-anonymity drops the $k$-fold replication requirement, it can be enforced through a broader range of mechanisms than $k$-anonymity. These include traditional methods used for $k$-anonymity (generalization, suppression) as well as permutation-based approaches (\emph{e.g.}, clustering records in groups of at least $k$ based on quasi-identifiers followed by random permutation of quasi-identifier values within each group \cite{probabilistic}, or the Anatomy mechanism~\cite{sun2009}). While generalization and suppression produce truthful but less detailed data, permutation-based methods yield perturbed data, but perfectly preserve the marginal distribution of every quasi-identifier attribute and, hence, its univariate statistics, such as means and variances. This can be advantageous for certain secondary analyses. Due to their randomness, permutation-based methods are also not reversible.  

On the other hand, since probabilistic $k$-anonymity is defined directly in terms of its privacy guarantee (a maximum reidentification probability of $1/k$ for any record), any (partially) reversible method --such as the above-mentioned {\em minimum} generalization attacked in~\cite{Cohen}-- would violate the definition. This makes the probabilistic $k$-anonymity model inherently robust against \emph{downcoding} attacks while keeping intact the guarantee intended by $k$-anonymity.

\subsection{Attribute disclosure}

A more obvious and dangerous weakness of $k$-anonymity is its inability to prevent attribute disclosure, as introduced in Section \ref{kanonymity}.
However, this can be fixed by applying $l$-diversity or $t$-closeness on top of $k$-anonymity.

Like in the case of $k$-anonymity, $l$-diversity offers a meaningful guarantee for any value of its parameter: for any $l>1$, unequivocal inferences of the confidential attribute value are no longer possible, 
and if a set of $l-1$ different values is augmented with an $l$-th different value, the intruder's uncertainty increases. 
This provides an \emph{ex ante} guarantee that is meaningful and holds for any value of $l$. However, due to the general formulation of $l$-diversity, which does not exactly prescribe the way to select the $l$ ``well-represented'' values, one should carefully consider the following
issues~\cite{tcloseness}:
\begin{itemize}
\item {\em Skewness.}  When the distribution of the confidential
attribute in the overall data set is strongly skewed, satisfying $l$-diversity
may be counterproductive as far as attribute disclosure is concerned.
Consider the case of a medical data set in which the confidential attribute
records the presence or absence of a given disease. Assume that 99\%
of the individuals are negative. Releasing a $k$-anonymous class with
50\% positives and 50\% negatives is optimal in terms of $l$-diversity
but is, indeed, potentially disclosive. After the release of such
data, each of the individuals in the $k$-anonymous class is found to have a 50\% probability of being positive in the listed disease,
while the prior probability (before the release) of being positive was only 1\%.
\item {\em Similarity.} Given $l' > l$, $l'$-diversity may be more disclosive than $l$-diversity if the 
semantic closeness of the $l'$ values is higher than that of the $l$ values. For example, a $k$-anonymous class containing $l'=4$ very similar income values involves more attribute disclosure than a class with $l=3$ more spread income values.
\end{itemize}

$t$-Closeness also protects against attribute disclosure, but unlike $l$-diversity, its guarantee becomes less and less meaningful as the value of $t$ increases. This loss of meaning for large parameter values is reminiscent of DP. In fact, $t$-closeness and DP are related: in \cite{connexio} it is shown that the information that an intruder obtains from accessing an $\exp(\epsilon)$-close microdata set satisfies $\epsilon$-DP for the confidential attribute. Indeed, DP guarantees that the knowledge gain obtained from the response to a query on an individual's confidential attribute is at most $\exp(\epsilon)$; that is, the distribution of the response must differ at most by a factor of $\exp(\epsilon)$ from the assumed prior knowledge, which is the marginal distribution of the confidential attribute. Therefore, just as the $\epsilon$-DP guarantee becomes meaningless for large values of $\epsilon$, so does $t$-closeness for large $t$.   

Moreover, $t$-closeness assumes that the marginal distributions of the confidential attributes are known. This makes sense when a marginal distribution
is not related to a specific individual. However, when one of the individuals in the data set is known
to have an outlying value (\emph{e.g.}, in a given economic sector the largest firm in the sector could be easily identifiable from the value of the confidential attribute "Turnover"), the marginal distribution
is itself disclosive. Also, if the entire data set shares the same (unknown) confidential value ({\em e.g.}, the data set is a sample of patients that are positive of a certain disease), even releasing a 0-close version will disclose the common value of the confidential attribute.

Both $l$-diversity and $t$-closeness can be enforced on top of $k$-anonymity --or, preferably, probabilistic $k$-anonymity-- by taking into account their constraints on the confidential attributes when forming the $k$-anonymous classes \cite{tkde}.

\subsection{Postprocessing immunity and composability}

Robustness against postprocessing requires that further processing of the protected output (without access to the original data) should still fulfill the privacy model. Since standard $k$-anonymity requires that each combination of values of the quasi-identifier attributes be shared by $k$ or more records, any postprocessing that breaks this rule, such as sampling or modifying individual records,  makes the output no longer $k$-anonymous.

The clash with postprocessing stems from the syntactic definition of standard $k$-anonymity, according to which 
the quasi-identifier values of any record must be replicated at least $k$
times in the $k$-anonymous data set.
However, {\em postprocessing does not affect the probability of reidentification}, since it is performed without access to the original data. Therefore, if one follows the suggestion above
to use a semantic
definition based on the risk of reidentification, postprocessing immunity is satisfied. In particular,
\emph{probabilistic $k$-anonymity satisfies postprocessing immunity}.  

Regarding composition, even though $k$-anonymity fulfills parallel composition (merging two $k$-anonymous disjoint data sets yields a $k$-anonymous data set), it does not provide a deterministic and smooth degradation of its privacy guarantee under sequential composition: merging two non-disjoint $k$-anonymous data sets may rapidly result in a 1-anonymous (\emph{i.e.}, unprotected) data set. 
Take a target
record belonging to the intersection of the two data sets and consider the intersection of the two
$k$-anonymous classes that contain the record (one from the first data set and the other from the second data set). If the two $k$-anonymous classes are different,
their intersection contains fewer than $k$ records but
is known to contain the target record; hence, $k$-anonymity is broken.
A specific consequence is that $k$-anonymity is not suitable for scenarios that involve incremental data releases on the same subjects.    

\section{Objections and mitigations to $\epsilon$-differential privacy}
\label{sec:dp}

The DP guarantee focuses mainly on protecting against \emph{membership 
disclosure}. 
In fact, the expression \eqref{eq:dp} 
refers to the intruder’s ability to infer membership, that is, the presence or absence of an individual's record in the data set. 
Protection against {\em identity disclosure}
and {\em attribute disclosure} are obtained as consequences:
\begin{itemize}
\item To make membership indistinguishable, all DP outputs must be perturbed so much that the values of attributes of original records remain protected; in other
words, the influence of the value of a particular attribute in a particular record on the DP outputs is negligible.
This protects against attribute disclosure.
\item If the presence or absence of any record cannot
be established and the record's attribute values cannot be
estimated from the DP outputs, the attacker cannot mount
a reidentification attack: reidentification needs a record
that can be attributed to someone.
\end{itemize}

This threefold guarantee is more robust than that of $k$-anonymity, but it often comes at the cost of unacceptable privacy-utility trade-offs, as we discuss below.

\subsection{DP guarantees do not hold for large $\epsilon$}

DP guarantees perfect indistinguishability only for $\epsilon=0$. More specifically, in this case, DP achieves Shannon's perfect secrecy~\cite{shannon1949communication},
regarding membership, identity, and attribute disclosures:
\begin{itemize}
\item {\em Perfect secrecy for membership and identity.} Let $T$ be a random variable that takes the value 1 if the target individual's record is a member of the original data set and the value 0 otherwise. Let $Y$ be the query output randomized with function $\kappa$ and $\epsilon=0$. Before seeing $Y$, the intruder's uncertainty on $T$ is $H(T)$, where $H$ stands for Shannon's entropy. After seeing $Y$, the intruder's
uncertainty on $T$ is $H(T|Y)$. Since expression \eqref{eq:dp} 
holds with equality for $\epsilon=0$, $Y$ tells nothing about the membership of any single record in the original data set, that is, $H(T|Y)=H(T)$, which is Shannon's perfect secrecy condition. 
Preventing membership inference effectively mitigates identity disclosure because reidentification presupposes membership (only members can be reidentified).
\item {\em Perfect secrecy for attribute values.} If we assume the value $R$ of the target record to be unknown to the intruder, the intruder's uncertainty
on $R$ is $H(R)$ before seeing the query answer $Y$.
After seeing $Y$, the intruder's uncertainty is $H(R|Y)$. When expression~\eqref{eq:dp}
holds with equality, $Y$ leaks nothing on the presence or absence of the 
target record, and hence it leaks nothing on $R$ so that $H(R|Y)=H(R)$. Thus, the value of $R$ remains perfectly secret.
\end{itemize}

The downside of perfect secrecy is that the query
answers have no connection whatsoever to the underlying original data, that is, the answers are completely random or, in other words, \emph{analytically useless}.
In practice, the inventors of DP also consider small $\epsilon$ values ({\em i.e.}, $\epsilon \leq 1$) ``safe'' \cite{dwork11}. 

Unfortunately, replacing $\epsilon=0$ with a small positive budget $\epsilon$
is not free from issues:
\begin{itemize}
\item On the utility side, 
a small budget $\epsilon$ is still incompatible with utility-preserving data releases, which are the use cases in most demand \cite{practice}.
\item On the privacy side, 
a small positive $\epsilon$ no longer ensures perfect indistinguishability or perfect protection
against disclosure of membership, identity, or confidential attributes. 
Strictly speaking,
the formal indistinguishability guarantee of DP 
fades away as $\epsilon$ grows.
Whatever protection is achieved under $\epsilon >0$ results from
the noise added to satisfy DP. Therefore, protection
against membership, identity, and attribute disclosure {\em must be empirically verified \underline{ex post}}. This is a major shortcoming because the main
attractive of using a privacy model is to dispense with such an empirical check. 
\end{itemize}

In this respect, many research works and most practical applications have employed unreasonably large $\epsilon$ (sometimes even more than 50 \cite{CensusPrivacyBudgetURL2023}) to keep their data useful enough \cite{Blanco2023, practice}. Although DP does not provide any real privacy guarantee in such loose implementations, most of them confidently state that their data are ``DP-protected'' and do not feel the need to experimentally evaluate {\em ex post} the actual privacy achieved \cite{practice}.

In an attempt to mitigate these issues and reach a better utility-privacy
trade-off, one may resort to DP relaxations. When introducing DP above,
we mentioned the most common DP relaxations, which allow smaller $\epsilon$ values to be achieved with less added noise, thereby improving utility preservation. However, this comes at the cost of a higher probability of violating the pure DP guarantee,  which adds further uncertainty about whether the \emph{ex ante} guarantees are satisfied. For example, if $(\epsilon,\delta)$-DP is used, there is a $\delta$ probability
that pure $\epsilon$-DP is not satisfied. Other relaxations of DP such as RDP or zCDP generally leak more private information than $\epsilon$-DP, and approach $\epsilon$-DP only asymptotically. 
Since these relaxations provide less privacy than pure $\epsilon$-DP, and the latter does not achieve perfect 
indistinguishability for $\epsilon>0$, the {\em ex ante} guarantees offered
by such relaxations are hardly sufficient to avoid the need
for {\em ex post} verification.

More interesting are the variants that deviate from DP in a deterministic way. 
\emph{Metric DP} \cite{metric} is in fact a generalization of DP that makes the guarantee of indistiguishability between two records $x$ and $x'$ not only inversely proportional to $\epsilon$ (as pure DP does), but also inversely proportional to the distance $d(x,x')$ between them. Closer records
are more indistinguishable from the metric DP output than more distant records. The intuition is that $\epsilon$ of DP is replaced by $\epsilon d(x,x`)$ in metric DP. 
Another option is \emph{individual DP} \cite{idp}, which requires indistinguishability only between records of the original data set and its corresponding neighbor data sets, instead of across any pair of neighbor data sets. As a result, it protects individuals with the {\em same} privacy guarantees as standard DP against identity/attribute disclosure, although the DP guarantees against membership disclosure are not preserved for groups of individuals. Both variants are deterministic relaxations of the DP guarantee that enable improved utility preservation for a given $\epsilon$ --or, conversely, comparable utility under smaller (stronger) $\epsilon$-- than pure DP. In fact, both have been designed to better cope with the needs of non-interactive data releases.

\subsection{Protecting membership incurs excessive perturbation and might be unnecessary}

As stated above, the DP definition specifically targets membership disclosure. However, since membership is a very general feature that concerns \emph{all} records in a data set, DP with small or even moderate $\epsilon$ requires adding random noise to most if not all records, regardless of how vulnerable they are to identity and/or attribute disclosure. This means that records with very common values, which are inherently protected against (identity and attribute) disclosure, get the same amount of noise as rarer records with more outlying values. The consequence is a severe degradation in the utility of DP outputs, even though protection against membership inference is very often
unnecessary, for example, when it does not implicitly entail attribute disclosure or when membership is already known --as in cases where the data set exhaustively represents a population--.

As in the previous section, one mitigation is to adopt alternative DP definitions that relax protection against membership disclosure. For example, 
\emph{individual DP} forgoes membership disclosure protection while preserving guarantees for individuals, allowing for significantly better utility when handling attributes with large or unbounded domains \cite{tdp}.

\subsection{Post-processing immunity and composability}

Post-processing immunity and composability are two widely acknowledged advantages of DP.
However, there are some nuances to consider when sequential composition is involved. 

If mechanisms \(\mathcal{A}_1\) and \(\mathcal{A}_2\) satisfy \(\epsilon_1\)-DP and \(\epsilon_2\)-DP, respectively, their combined application on non-disjoint data satisfies $(\epsilon_1 + \epsilon_2)\text{-DP}$. Although this looks like a smooth degradation of the privacy guarantees, we should realize that:
\begin{itemize}
\item The indistinguishability guarantee of {\em decreases exponentially} with $\epsilon$. Therefore, adding two budgets $\epsilon_1 + \epsilon_2$ involves an exponential degradation in privacy with respect to the situation prior to sequential composition. 
\item Under sequential composition --which applies to all attributes and non-independent records in data releases, or to successive data releases or data collection on the same individuals--, the effective $\epsilon$ grows quickly. This issue has been observed in many practical applications, where sequential composition led to two- or even three-digit values of $\epsilon$, despite initially setting seemingly 'safe' values of $\epsilon \leq 1$ \cite{practice}.
\end{itemize}

Thus, while the sequential composition of DP offers a straightforward way to quantify the effective $\epsilon$ when dealing with non-disjoint data, it does not solve the challenge of ensuring meaningful privacy in scenarios such as successive data releases or continuous data collection.

\section{Conclusions}
\label{sec:conclusion}

Privacy models are an attractive option for data anonymization, because they promise to free the data protector from the need to perform costly, tedious, context-specific, and error-prone {\em ex post} assessments of the extant risk of disclosure. 
In this paper, we have explored to what extent this promise is upheld by the two main families of privacy models, namely the $k$-anonymity family and the differential privacy family.
Our findings are that {\em both} families face objections that must be mitigated for their {\em ex ante} guarantees to be effective. This contradicts the common belief that $k$-anonymity is more obsolete than DP and too limited for practical use, while all that is necessary to protect privacy is to enforce DP and ``never look back'' \cite{dwork2007}: Both families have their own shortcomings, but also complementary strengths.

The syntactic definition of $k$-anonymity poses formal problems because, when enforced through minimum generalization, it may allow recovering the original data from the $k$-anonymous data. On the other hand, it does not satisfy postprocessing immunity. However, {\em taking a semantic definition of $k$-anonymity focused
on limiting re-identification risk to $1/k$ is a sound
solution} that fixes both issues. 
As for the well-known lack of protection against attribute disclosure, it is meaningfully fixed by extensions such as $l$-diversity or $t$-closeness. Overall, $k$ anonymity provides a robust solution to protect data releases and, through its targeted masking, can achieve a more favorable privacy–utility trade-off compared to DP. 

In the case of DP, strict indistinguishability is satisfied only for $\epsilon=0$. For positive $\epsilon$, indistinguishability is not perfect. Furthermore, small positive $\epsilon$ values are likely to result in an unacceptable loss of utility, whereas larger $\epsilon$ values offer an indistinguishability guarantee so degraded that it is mostly void. This is mitigated through the use of relaxations, which have been widely adopted in practical applications \cite{practice}. However, those relaxations that entail a probability of the DP condition not holding further degrade the {\em ex ante} guarantee and can hardly relieve the data protector of the need to conduct an {\em ex post}
evaluation of the disclosure risk. This basically cancels the attractiveness
of using DP as a privacy model. Variants of DP that relax the guarantee in a deterministic way are more reassuring, while still providing significant utility gains. Anyway, selecting an appropriate DP relaxation --among the dozens proposed in the literature \cite{pejo2020sok}-- remains a major challenge for practitioners, as it must balance meaningful parameterization with task-specific utility preservation. Nevertheless, DP is better suited to the interactive applications for which it was originally designed --applications that, notably, have been the only ones to successfully reconcile reasonable utility with meaningful privacy parameters in practice \cite{practice}. 

$k$-Anonymity does not offer composability when applied to several
overlapping data sets. In contrast, DP offers attractive composition properties, although indistinguishability exponentially weakens at each sequential composition. Admittedly, no other privacy model in
the literature performs better than DP in the case of sequential composition.
This suggests that current privacy models are not very well suited
for protecting incremental data releases or data streams.
It is an open problem to find privacy models whose privacy guarantee
is meaningful and degrades more gracefully after sequential
composition \cite{practice}.

\section*{Acknowledgements}
Partial support to this work has been received from the Government of Catalonia (ICREA Acad\`emia Prizes to J. Domingo-Ferrer and to D. S\'anchez, and grant 2021SGR-00115), 
MCIN/AEI under grant PID2024-157271NB-I00 ``CLEARING-IT'', and 
the EU's NextGenerationEU/PRTR via INCIBE (project ``HERMES'' and INCIBE-URV cybersecurity chair).

\bibliographystyle{abbrv} 
\bibliography{source}

\end{document}